\documentclass[final,1p,times,twocolumn]{elsarticle}
\usepackage{graphicx,amssymb}
\journal{Physics Letters B}
\begin{document}

\newcommand{\be}{\begin{equation}}
\newcommand{\ee}{\end{equation}}
\newcommand{\bq}{\begin{eqnarray}}
\newcommand{\eq}{\end{eqnarray}}
\newcommand{\bsq}{\begin{subequations}}
\newcommand{\esq}{\end{subequations}}
\newcommand{\bc}{\begin{center}}
\newcommand{\ec}{\end{center}}

\begin{frontmatter}

\title{On the Stability of Fundamental Couplings in the Galaxy}
\author[inst1,inst2]{S. M. Jo\~ao}\ead{up201206827@fc.up.pt}
\author[inst1,inst3]{C. J. A. P. Martins\corref{cor1}}\ead{Carlos.Martins@astro.up.pt}
\author[inst1,inst2]{I. S. A. B. Mota}\ead{up200908588@fc.up.pt}
\author[inst1,inst2]{P. M. T. Vianez}\ead{up201201345@fc.up.pt}
\address[inst1]{Centro de Astrof\'{\i}sica, Universidade do Porto, Rua das Estrelas, 4150-762 Porto, Portugal}
\address[inst2]{Faculdade de Ci\^encias, Universidade do Porto, Rua do Campo Alegre, 4150-007 Porto, Portugal}
\address[inst3]{Instituto de Astrof\'{\i}sica e Ci\^encias do Espa\c co, CAUP, Rua das Estrelas, 4150-762 Porto, Portugal}
\cortext[cor1]{Corresponding author}

\begin{abstract}
Astrophysical tests of the stability of Nature's fundamental couplings are a key probe of the standard paradigms in fundamental physics and cosmology. In this report we discuss updated constraints on the stability of the fine-structure constant $\alpha$ and the proton-to-electron mass ratio $\mu=m_p/m_e$ within the Galaxy. We revisit and improve upon the analysis by Truppe {\it et al.} \cite{Truppe} by allowing for the possibility of simultaneous variations of both couplings and also by combining them with the recent measurements by Levshakov {\it et al.} \cite{Levshakov}. By considering representative unification scenarios we find no evidence for variations of $\alpha$ at the 0.4 ppm level, and of $\mu$ at the 0.6 ppm level; if one uses the \cite{Levshakov} bound on $\mu$ as a prior, the $\alpha$ bound is improved to 0.1 ppm. We also highlight how these measurements can constrain (and discriminate among) several fundamental physics paradigms.
\end{abstract}


\end{frontmatter}

\section{Introduction}

Nature's dimensionless fundamental couplings are among the deepest mysteries of modern physics: it is clear that they play a crucial role in physical theories, and yet we have no 'theory of constants' that describes what this role is. Three rather different views on the subject are discussed in \cite{Duff}, and a broader overview of the subject can be found in Uzan's review \cite{Uzan}. At a phenomenological level it is well known that fundamental couplings {\it run} with energy, and in many extensions of the standard model they will also {\it roll} in time and {\it ramble} in space ({\it i.e.}, they will depend on the local environment). The class of theories with additional spacetime dimensions, such as string theory, is the most obvious example.

An unambiguous detection of varying dimensionless fundamental couplings will be revolutionary: it will establish that the Einstein equivalence principle is violated and that there is a fifth force of nature. We refer the interested reader to \cite{Uzan} as well as to the recent Equivalence Principle overview by Damour \cite{Damour} for detailed discussions of these points. Nevertheless, improved null results are almost as important. Naively, the natural scale for the cosmological evolution of one of these couplings (if one assumes the simplest paradigm, in which it is driven by a scalar field) would be the Hubble time, and we would therefore expect a drift rate of the order of $10^{-10} {\rm yr}^{-1}$. However, local tests with atomic clocks \cite{Rosenband} restrict any such drift to be at least six orders of magnitude weaker, and thereby rule out may otherwise viable models. This explains why tests of the stability of nature's fundamental couplings are among the key drivers for the next generation of ESO and ESA facilities. Additionally, these tests have important implications for the enigma of dark energy, as discussed in \cite{Amendola}.

Evidence for spacetime variations of the fine-structure constant $\alpha$, in the redshift range $z\sim1-4$ and at the few parts per million level has been provided by \cite{Webb}. An ongoing Large Program at ESO's Very Large Telescope is independently testing these results, and the first results of this effort have recently been reported by \cite{Molaro}. Given the limitations of current optical/UV spectrographs, a definitive answer may have to wait for a forthcoming generation of high-resolution ultra-stable spectrographs, such as ESPRESSO and ELT-HIRES \citep{Espresso,Hires}, both of which include improving these measurements among their key science/design drivers \cite{Martins}. Radio/microwave measurements of these couplings can also be performed. While they are typically limited to lower redshifts than their optical/UV counterparts, they sensitivity is competitive. A meta-analysis of the various recent early universe measurements can be found in \cite{Ferreira}.

Another advantage of the radio/microwave band for our purposes is that they allow measurements within the Galaxy (effectively at $z=0$) which provide tests of possible environmental dependencies. Recently \cite{Truppe} provided improved constraints on the stability of $\alpha$ and also the proton-to-electron mass ratio, $\mu=m_p/m_e$. However, a constraint for each of these was derived on the assumption that the other does not vary. The authors of \cite{Truppe} explicitly recognize in their own paper that this is a weakness of their analysis. Far from being just a harmless simplification, from a theoretical point of view this is an unnatural assumption which (as we will show) can lead to seemingly tight but misleading bounds. In this work we overcome this limitation, and also combine their dataset with the recent direct measurement of $\mu$ by \cite{Levshakov}, thus improving on an analysis by \cite{LevF10}. We note that in this note we define $\mu=m_p/m_e$, in accordance with the cosmology/particle physics standard practice, while \cite{Truppe,Levshakov} use $\mu=m_e/m_p$ (in accordance with atomic physics conventions).

\section{Analysis and Results}

Recently \cite{Truppe} derived a set of constraints on the stability of fundamental couplings by comparing laboratory and astrophysical measurements of selected microwave transitions in CH and OH molecules. The rest frequency emitted by the astrophysical source and the laboratory frequency are related by
\begin{equation}
\omega_{\rm ast}=\omega_{\rm lab}\left[1+K_\alpha\frac{\Delta\alpha}{\alpha}+K_\mu\frac{\Delta\mu}{\mu}\right]\,,
\end{equation}
where $K_\alpha$ and $K_\mu$ are the sensitivity coefficients for the transition in question, quantifying how much it is affected by a given amount of change in $\alpha$ and $\mu$.
The precise sensitivity coefficients for the relevant CH and OH transitions, which are typically of order unity, can be found in \cite{Kozlov,Denijs}. With this information \cite{Truppe} separately obtain bounds for $\Delta \alpha / \alpha$ and $\Delta \mu / \mu$, respectively assuming that the other coupling does not vary. In this case the fractional variation of $\alpha$ can be obtained by comparing two different transitions in the same system, and will be given by
\begin{equation}
\frac{\Delta \alpha}{\alpha} = \frac{1}{K_{\alpha_2}-K_{\alpha_1}}\frac{\Delta v'_{12}}{c}\,,
\end{equation}
with an analogous expression for $\mu$. Here $\Delta v'_{12}$ is a suitably corrected difference between the measured velocities of the two transitions in question. Table \ref{tableTruppe} summarizes their results. Specifically, they find the following weighted mean average of the results for the five different sources (displayed with one-sigma uncertainties)
\begin{equation}\label{truppe1}
\left(\frac{\Delta \alpha}{\alpha}\right)_{\rm Truppe} = (0.32\pm1.08)\times10^{-7}
\end{equation}
\begin{equation}\label{truppe2}
\left(\frac{\Delta \mu}{\mu}\right)_{\rm Truppe} = (0.68\pm2.23)\times10^{-7}\,.
\end{equation}

%----------------------------------------------------------- S_vib
\begin{table}
\begin{center}                         % used for centering table
\begin{tabular}{c c c c}        % centered columns (4 columns)
\hline                 % inserts double horizontal lines
Source & $\Delta v'_{12}$ \ (km s$^{-1}) $ & $\Delta \alpha / \alpha \  (10^{-7})$ & $\Delta \mu / \mu \  (10^{-7})$  \\    % table heading 
\hline                        % inserts single horizontal line
   G111.7-2.1 & $-0.08\pm0.11$ & $+1.5\pm2.0$ & $+3.1\pm4.1$ \\      % inserting body of the table
  G265.1+1.5  & $ +0.04\pm 0.16$ & $-0.9\pm 3.1$    & $-1.9 \pm6.4$ \\
  G174.3-13.4 & $-0.02\pm0.19$ & $+0.6 \pm3.6$     & $+1.2\pm7.4$ \\
  G6.0+36.7 & $-0.12\pm0.13$ & $+2.3\pm2.4$    & $+4.8\pm5.0$ \\
  G49.5-0.4  & $-0.48\pm0.55$ & $-1.8\pm2.0$    & $-3.6\pm4.1$ \\ 
\hline                                   %inserts single line
\end{tabular}
\caption{\label{tableTruppe}Data from the five interstellar sources used by (and reproduced from) \protect\cite{Truppe}. Both the velocity differences and the fractional variations are given with one-sigma uncertainties. Note that our definition of $\mu$ differs from that of \protect\cite{Truppe}.}
\end{center}
\end{table}
%----------------------------------------------------------- S_vib

However, assuming that one constant is fixed while the other varies has no generic theoretical motivation. Instead, one generically expects that the two couplings will vary simultaneously, with the relative size of the variations being highly model-dependent. For example, in a broad class of unification scenarios, discussed in \cite{Coc} and recently tested against extragalactic measurements in \cite{Ferreira}, the two variations are related by
\begin{equation}
\frac{\Delta \mu}{\mu} = [0.8R-0.3(1+S)]\frac{\Delta \alpha}{\alpha}\,,
\end{equation}
where $R$ and $S$ are true dimensionless fundamental couplings (meaning that they are spacetime-invariant), with the former being related to Quantum Chromodynamics and the latter to the Electroweak sector of the underlying theory. Thus different models will be characterized by different values of $R$ and $S$. In particular, whether the two variations have the same or opposite signs is model-dependent. Importantly, note that the fact that these parameters are assumed to be universal makes them ideal for comparing measurements obtained in different contexts: for example, bounds on $R$ and $S$ obtained in local laboratory tests should also apply to astrophysical systems.

%----------------------------------------------------------- S_vib
\begin{figure}
\begin{center}
\includegraphics[width=2.7in]{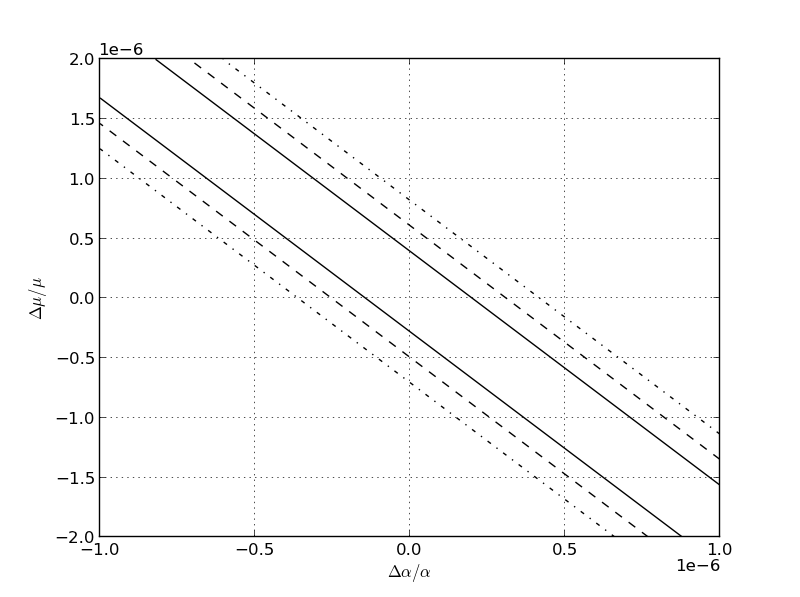}
\caption{\label{figAlphaMu}Constraints on the $\alpha-\mu$ parameter space, obtained from the data in Table \protect\ref{tableTruppe} while allowing for generic simultaneous variations of both couplings. One, two and three sigma constraints are respectively indicated by the solid, dashed and dash-dotted lines.}
\end{center}
\end{figure}
%----------------------------------------------------------- S_vib

Clearly the molecular transitions being used are sensitive to changes of both $\alpha$ and $\mu$, and inspection of the sensitivity coefficients shows that the sensitivity to the former is twice that of the latter, meaning that the result of \cite{Truppe} is actually a constraint on the product of both, namely
\begin{equation}\label{boundf}
\frac{\Delta (\alpha^2\mu)}{(\alpha^2\mu)} = (0.68\pm2.23)\times10^{-7}\,.
\end{equation}
(Strictly speaking the ratio of the two sensitivities is 2.01 according to the calculations of \cite{Kozlov,Denijs}, but in what follows we will simply assume it to be 2; this nominal one percent difference is clearly negligible in comparison with other theoretical and observational uncertainties.)

Using standard least squares techniques we can constrain the $\alpha-\mu$ parameter space with the above data, while allowing for generic simultaneous variations of both couplings. The results of this analysis are shown in Fig. \ref{figAlphaMu}, which makes the presence of this degeneracy obvious. This shows that the bounds given by Eqs. (\ref{truppe1}--\ref{truppe2}) are misleading: there's an infinite number of models (i.e., choices of $R$ and $S$) that can be consistent with Eq. (\ref{boundf}) but nevertheless have $\alpha$ and $\mu$ variations larger than those given by Eqs. (\ref{truppe1}--\ref{truppe2}).

There are, however, ways to break this degeneracy. A simple, model-independent one is to use as external prior an independent measurement of one of the two couplings. This was done in \cite{LevF10}, upon which we can improve by using the recent measurement of the proton-to-electron mass ratio in the Galactic plane by \cite{Levshakov}, which with our convention for $\mu$ is
\begin{equation}
\left(\frac{\Delta \mu}{\mu}\right)_{\rm Levshakov} = (-0.03\pm0.06)\times10^{-7}\,.
\end{equation}
Substituting this in Eq. (\ref{boundf}) then leads to the following bound for the fine-structure constant
\begin{equation}
\frac{\Delta \alpha}{\alpha} = (0.36\pm1.12)\times10^{-7}\,.
\end{equation}
Although nominally this is consistent with the result of \cite{Truppe} (and indeed very close to it, since the prior on $\mu$ has a very small statistical uncertainty compared to that of $\alpha^2\mu$), we emphasize that physically it is a much more robust bound.

%----------------------------------------------------------- S_vib
\begin{table}
\begin{center}                         % used for centering table
\begin{tabular}{c c | c c}        % centered columns (4 columns)
\hline                 % inserts double horizontal lines
Assumption & References & $\Delta \alpha / \alpha \  (10^{-7})$ & $\Delta \mu / \mu \  (10^{-7})$  \\    % table heading 
\hline                        % inserts single horizontal line
Other fixed & \cite{Truppe} only & {\it +0.32$\pm$ 1.08} & {\it +0.68$\pm$ 2.23} \\      % inserting body of the table
\hline
Unification scenario & \cite{Truppe} $+$ \cite{Coc}  & $-0.04 \pm0.13$     & $+0.76\pm2.48$ \\
Dilaton-type model & \cite{Truppe} $+$ \cite{Nakashima}  & $+0.01\pm0.03$    & $+0.66\pm2.18$ \\
Atomic clocks & \cite{Truppe} $+$ \cite{Clocks}  & $+1.36\pm4.46$    & $-2.04\pm6.69$ \\ 
\hline
Direct $\mu$ measurement & \cite{Truppe} $+$ \cite{Levshakov} & $+0.36\pm1.12$ & {\it -0.03$\pm$ 0.06} \\
\hline                                   %inserts single line
\end{tabular}
\caption{\label{models}Comparison of the constraints on variations of $\alpha$ and $\mu$ from the data in Table \protect\ref{tableTruppe}, under several different assumptions. The inferred fractional variations are given with one-sigma uncertainties. Values in italics were obtained in the papers listed in the second column; the others are the result of the present work.}
\end{center}
\end{table}
%----------------------------------------------------------- S_vib

Alternatively one may focus on particular models, which will provide specific values of the unification parameters $R$ and $S$. As discussed by \cite{Coc}, current (possibly naive) expectations regarding unification scenarios may suggest that typical values would be
\begin{equation}
R\sim36\,,\quad S\sim160\,.
\end{equation}
Nevertheless, it's important to realize that these values are highly model-dependent, and they can vary widely among different classes of models. As an example, in the dilaton-type model whose variations of fundamental couplings have been studied by \cite{Nakashima} one has
\begin{equation}
R\sim109.4\,,\quad S\sim0\,.
\end{equation}
Finally, \cite{Clocks} provide constraints on the local drift of $\alpha$ and $\mu$ (as well as that of the proton gyromagnetic ratio) from laboratory comparisons among atomic clocks with different sensitivities to these couplings, and translate these into constraints on $R$ and $S$. One finds a degeneracy implying that one can only constrain the combination
\begin{equation}
(1+S)-2.7R=-5\pm15\,;
\end{equation}
here for simplicity we will simply assume the best-fit value. Note that this last case, although not a purely theoretical prior (as it relies on atomic clock data), is not completely model-independent since a class of unification scenarios is being assumed.

In Table \ref{models} we compare the bounds on the fractional variations of $\alpha$ and $\mu$ obtained by \cite{Truppe} under the assumption that the other coupling is fixed with those obtained assuming that the two are related in the three ways discussed above (without including the direct measurement of $\mu$), as well as with those obtained with the \cite{Levshakov} measurement as prior. Our goal here is not to obtain tighter constraints than \cite{Truppe}, but to provide a more robust analysis and also to highlight the fact that derived constraints are highly model-dependent. Moreover, in as much as the three model assumptions on $R$ and $S$ are representative of a vast parameter space, it's clear that assuming that the other coupling is fixed can lead to erroneously tight constraints.

A more robust procedure is therefore to combine different datasets or to use external observational priors. A possible {\it caveat} here is that in models where the couplings depend on the environment (specifically, the local density) combining measurements from different environments still requires assumptions on the underlying model. Thus rather than combining $\alpha$ and $\mu$ datasets directly one should instead translate them into bounds on the $R$--$S$ parameter space, since these are expected to be spacetime-invariant.

\section{Discussion and Conclusions}

We have revisited recent astrophysical tests of the stability of fundamental couplings in the Galaxy and assessed them in a theoretical context. The main point of our brief analysis is to emphasize that, when measuring quantities that depend on a combination of several couplings, inferring constraints on one of the couplings by assuming that the others do not vary is not only theoretically unjustified but may well lead to unrealistic constraints, in the sense that in at least part of the range of models the derived constraints will be considerably weaker. By considering some representative unification scenarios we find no evidence for variations of $\alpha$ at the 0.4 ppm level, and of $\mu$ at the 0.6 ppm level; if one uses the \cite{Levshakov} bound on $\mu$ as a prior, the $\alpha$ bound is improved to 0.1 ppm. The sensitivity of these constraints is thus comparable to those obtained from the Oklo natural nuclear reactor \cite{Oklo}, if the latter are expressed as constraints on $\alpha$.

We note that this assumption (constraining one coupling by fixing the others) is as prevalent in the literature and unrealistic as that of a constant drift rate (in other words, assuming that measurements at non-zero redshift can be related to local measurements by assuming a linear time variation of the relevant coupling). While naively they may seem harmless and conservative---and thus reasonable or conservative approximations---this is in fact not the case. There are no realistic models for which they will hold, and indeed (as we have explicitly shown) one can easily find examples where the putative constraints derived under these assumptions are violated.

While astrophysical measurements that are simultaneously sensitive to a combination of various fundamental couplings, such as $\alpha$, $\mu$ and the proton gyromagnetic ratio, can play an important role in the quest for new physics beyond the standard cosmology and particle physics paradigms, the optimal way to use them is in combination with direct measurements of $\alpha$ or $\mu$, as we illustrated above. Alternatively, if several of these measurements are sensitive to different combinations of the relevant constants, a joint analysis of data will reduce or break degeneracies between parameters and lead to robust constraints on individual couplings. This has been done, for example, by \cite{Clocks} using atomic clock data and by \cite{Ferreira} with extra-galactic measurements.

Indeed these joint measurements, for which the number of known targets is relatively large (both within the Galaxy and outside it) may be crucial for the future of the field. Ultimately, as discussed in \cite{Amendola}, one would like to map the behavior of $\alpha$ and $\mu$ in the entire range from $z=0$ to deep in the matter era (say $z\sim5$ or even beyond), in order to constrain the dynamics of putative scalar fields. In some redshifts, targets that may provide stringent constraints on $\alpha$ and $\mu$ are extremely scarce, and in those cases joint measurements can profitably be used and included in a more extensive analysis.

\section*{Acknowledgements}
This work was done in the context of project PTDC/FIS/111725/2009 (FCT, Portugal). CJM is also supported by an FCT Research Professorship, contract reference IF/00064/2012, funded by FCT/MCTES (Portugal) and POPH/FSE (EC).

\bibliographystyle{model1-num-names}
\bibliography{galaxy}

\end{document}